# Dielectric and gate metal engineering for threshold voltage modulation in enhancement mode monolayer MoS$_2$ field effect transistors

Lixin Liu[1], Han Yan[1], Leyi Loh[1], Kamal Kumar Paul[1], Soumya Sarkar[1], Deepnarayan Biswas[2], Tien-Lin Lee[2], Takashi Taniguchi[3], Kenji Watanabe[4], Manish Chhowalla[1]* & Yan Wang[1]*

**Affiliations:**

[1]*Department of Materials Science & Metallurgy, University of Cambridge, 27 Charles Babbage Road, Cambridge CB3 0FS, UK*

[2]*Diamond Light Source, Harwell Science and Innovation Campus, Didcot OX11 0DE, United Kingdom*

[3]*Research Center for Materials Nanoarchitectonics, National Institute for Materials Science, 1-1 Namiki Tsukuba Ibaraki 305-0044, Japan.*

[4]*Research Center for Electronic and Optical Materials, National Institute for Materials Science, Tsukuba, Japan.*

*Correspondence to: yw472@cam.ac.uk, mc209@cam.ac.uk

**Abstract:** Excellent gate electrostatics in field effect transistors (FETs) based on two-dimensional transition metal dichalcogenide (2D TMD) channels can dramatically decrease static power dissipation. Energy efficient FETs operate in enhancement mode with small and positive threshold voltage ($V_{th}$) for n-type devices. However, most state-of-the-art FETs based on monolayer MoS$_2$ channel operate in depletion mode with negative $V_{th}$ due to doping from the underlying dielectric substrate. In this work, we identify key properties of the semiconductor/dielectric interface (MoS$_2$ on industrially relevant high dielectric constant (*k*) HfO$_2$, ZrO$_2$ and hBN for reference) responsible for realizing enhancement-mode operation of 2D MoS$_2$ channel FETs. We find that hBN and ZrO$_2$ dielectric substrates provide low defect interfaces with MoS$_2$ that enables effective modulation of the $V_{th}$ using gate metals of different work functions (WFs). We use photoluminescence (PL) and synchrotron X-ray photoelectron spectroscopy (XPS) measurements to investigate doping levels in monolayer $MoS_2$ on different dielectrics with different WF gate metals. We complement the FET and spectroscopic measurements with capacitance-voltage analysis on dielectrics with varying thicknesses, which confirm that $V_{th}$ modulation in $ZrO_2$ devices is correlated with WF of the gate metals – in contrast with $HfO_2$ devices that exhibit signatures of $V_{th}$ pinning induced by oxide/interface defect states. Finally, we demonstrate FETs using a 2D $MoS_2$ channel and a 6 nm of $ZrO_2$ dielectric, achieving a subthreshold swing of 87 mV dec$^{-1}$ and a threshold voltage of 0.1 V. Our results offer insights into the role of dielectric/semiconductor interface in 2D $MoS_2$ based FETs for realizing enhancement mode FETs and highlight the potential of $ZrO_2$ as a scalable high-$k$ dielectric.

**Introduction:**

Atomically thin semiconductors, particularly transition metal dichalcogenides (TMDs) such as monolayer $MoS_2$, have emerged as promising candidates for next-generation electronics.[1-5] While field effect transistors (FETs) based on TMDs hold promise and complementary FETs, 32-bit microprocessor, gigahertz circuits, and microwave transmitters have been demonstrated [6-8], most reported n-type devices operate in depletion mode (threshold voltage, $V_{th}$ < 0), exhibiting considerable channel current (typically > nA/μm) at zero gate bias.[8-14] FETs that operate in enhancement-mode ($V_{th}$ > 0) reduce the power consumption by orders of magnitude.[15-16]

Despite the importance of realizing stable and reproducible $V_{th}$ for FETs based on atomically thin TMDs, studies on underlying mechanisms governing $V_{th}$ remain limited. In silicon metal oxide semiconductor FETs (MOSFETs) beyond the 45 nm node, precise $V_{th}$ is achieved via gate metal WF. That is, gate metals on top of the high $k$ dielectric are selected to align near the conduction or valence band edges of silicon to achieve low and symmetric $V_{th}$ values for nMOS and pMOS operation.[17, 18] This strategy has been successfully implemented in FinFETs[19] and gate-all-around (GAA) FETs, where matched $V_{th}$ values (~ 0.35 V) have been demonstrated.[20] Interface dipole engineering has also been widely studied to develop multiple $V_{th}$ for Si MOSFETs. [21,22] In emerging technologies such as carbon nanotube FETs, gate WF tuning has achieved $V_{th}$ shifts of ~ 0.5 V without introducing dopants.[23]

For $MoS_2$ FETs, approaches such as solvent doping and the introduction of interface seed layer have been explored to tune the $V_{th}$ of $MoS_2$ FETs.[24-26] While gate WF modulation remains a promising strategy for $V_{th}$ control in FETs based on monolayer $MoS_2$, experimental reports often show divergent operation modes when using the same metal gate metals.[8,9,12,27-30] These discrepancies highlight that the fundamental mechanism governing $V_{th}$ control remains poorly understood. As a result, achieving controlled $V_{th}$ is a key challenge for enabling enhancement-mode operation in monolayer TMD FETs.

Here, we systematically examine how different dielectrics (hBN, $HfO_2$ and $ZrO_2$) with different WF gate metals (Al, Au, and Pt) influence doping of 2D $MoS_2$ semiconducting channel. Photoluminescence (PL) and X-ray photoelectron spectroscopy (XPS) measurements were used to study degree of doping and binding energy shift (indicative of Fermi level shift), respectively, in the $MoS_2$ channel with different gate metals. We found that metals with different WFs coupled with $ZrO_2$ and hBN dielectrics that form a clean semiconductor/dielectric interface can modulate the carrier concentration in the $MoS_2$ channel so that the $V_{th}$ can be tuned from negative to positive when going from Al gate metal (WF ~ 4.1 eV) to Pt (WF ~ 5.2 eV) [31, 32]. In contrast, unintentional electron doping at the interface between $HfO_2$ and monolayer $MoS_2$ prevents modulation of $V_{th}$, which remains fixed around −0.5 V regardless of gate metal.

**Results and discussion:**

To investigate the influence of gate metal on devices based on monolayer $MoS_2$, we fabricated back-gated FETs incorporating metal gates with variable WFs. The device configuration is shown in **Figure 1a**. As illustrated in **Figures 1b** and **1c**, the band bending in metal-insulator-semiconductor (MIS) junctions is modulated by the WF difference between the gate and semiconductor. A low WF metal induces electron accumulation at the semiconductor interface under equilibrium, which leads to negative shift of $V_{th}$. Conversely, high WF metals deplete electrons and shift $V_{th}$ to positive values. In principle, WF-induced band bending and carrier redistribution should be even more pronounced in monolayer channels due to their negligible screening thickness. However, atomically thin $MoS_2$ is extremely sensitive to its dielectric environment[2, 33], hence the dielectric/semiconductor interface quality critically impacts the effectiveness of gate WF modulation of carriers in the channel. To clarify these effects, we compare FETs with different dielectrics.

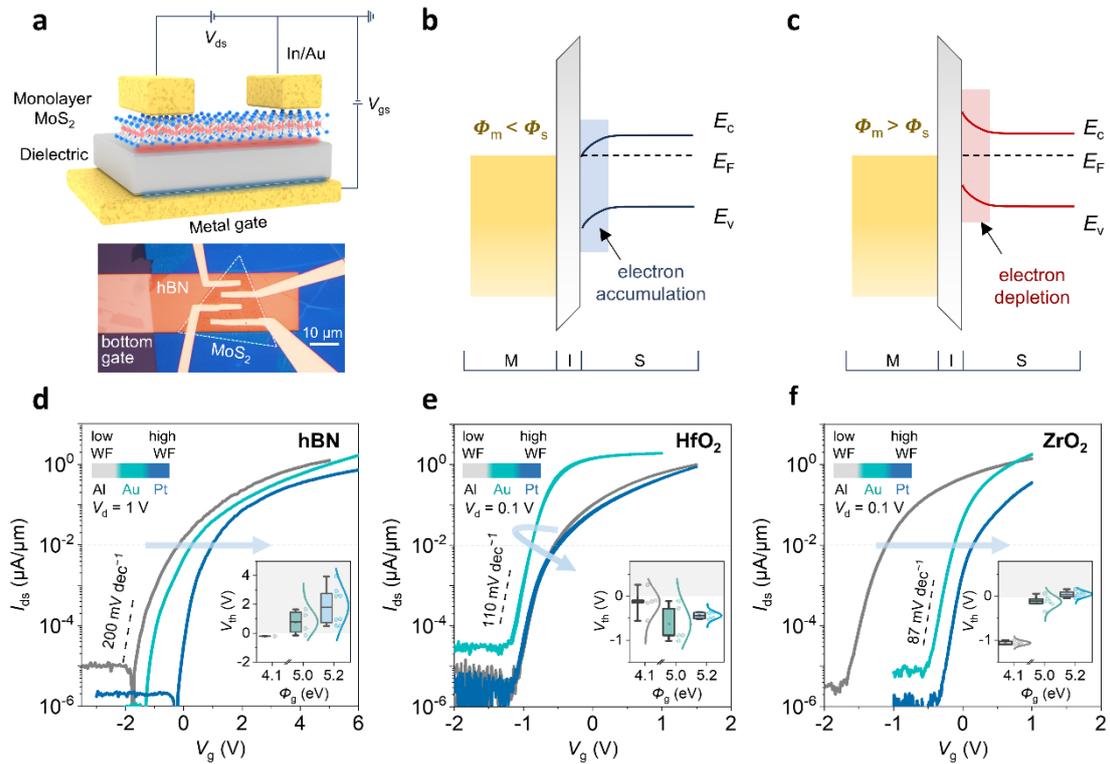

**Figure 1: Performance of monolayer $MoS_2$ FETs with gate work function engineering. a**, Schematic illustration and optical microscopy (OM) image of the fabricated monolayer $MoS_2$ FETs. **b**, Energy band alignment of MIS junction with a low WF metal gate. **c**, Energy band alignment of MIS junction with a high WF metal gate. **d-f**, Transfer characteristics of devices incorporating different gate metals on **d**, 12 nm of hBN, **e**, 6 nm of $HfO_2$, and **f**, 6 nm of $ZrO_2$. The inserted figures show the extracted $V_{th}$ as a function of gate metal work function for multiple devices with different dielectrics.

The representative transfer characteristics for devices based on monolayer $MoS_2$ with 12 nm of hBN, 6 nm of $HfO_2$, and 6 nm of $ZrO_2$ dielectrics are shown in **Figures 1d-f**. We use hexagonal boron nitride (hBN) for comparison because it provides an atomically flat, inert interface that preserves the intrinsic properties of $MoS_2$.[34, 35] We observe that as the underlying gate metal is varied from low to high WFs, the transfer curves shift systematically toward positive gate voltages, consistent with the expected modulation of $V_{th}$. The hBN dielectric FETs require relatively large gate voltages (typically > 6 V) to achieve on state because of the low dielectric constant of hBN, which restricts charge induction efficiency in the channel (details in **Figure S1**). In contrast, high-*k* dielectrics such as $HfO_2$ and $ZrO_2$ greatly enhance electrostatic coupling, enabling relatively low-voltage operation. The device characteristics in **Figure 1e** show that the transfer curves for $HfO_2$-based devices nearly overlap, indicating the $V_{th}$ is independent of back gate metal. In contrast, $V_{th}$ shifts towards positive value with increasing gate metal work function with $ZrO_2$ dielectric, following the same trend as hBN FETs but operating at lower biases. Detailed statistical analyses of $V_{th}$ for multiple devices with different gate metals are shown in the inserted figures (**Figure 1d-f**). $V_{th}$ exhibits a linear dependence with metal gate work function when hBN is used as the dielectric. However, hBN devices show large variations from device to device because the viscoelastic transfer of mechanically exfoliated hBN for device fabrication introduces air gaps and surface residue. $ZrO_2$-based FETs also exhibit a monotonic increase in $V_{th}$ with increasing gate WF, showing approximately a 1.0 V shift in $V_{th}$ using low and high WF metals. Pt-gated $ZrO_2$ devices operate in enhancement mode with an average $V_{th}$ of ~0.1 V. FETs using $ZrO_2$ as the dielectric were reproducible in their performance and trend. In contrast, $HfO_2$-based devices exhibit a nearly fixed, negative $V_{th}$ of approximately −0.5 V, independent of the gate metal work function. Furthermore, as indicated in **Figure 1d-f** and **Figures S2** and **S3**, $ZrO_2$-based FETs exhibit the lowest subthreshold swing (SS) values. The favourable electrical performance such as sub 1 V operating voltage, enhancement-mode operation (average $V_{th}$ ~0.1 V), and low SS (~ 87 mV dec$^{-1}$), demonstrate that $ZrO_2$-based $MoS_2$ FETs with Pt gates closely align with the performance requirements for state-of-the-art low-power FETs.[3]

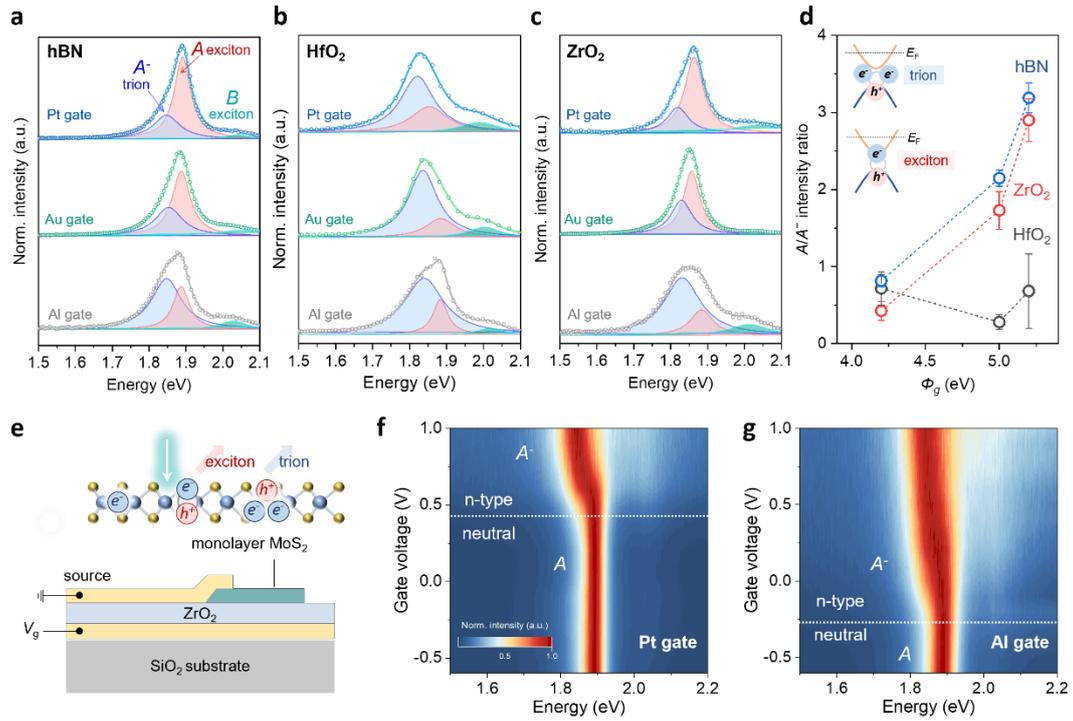

**Figure 2: PL analysis of monolayer $MoS_2$ with dielectric and gate metal engineering. a–c**, PL spectra of monolayer $MoS_2$ on gate metals with different WFs, measured on **a**, hBN, **b**, $HfO_2$, and **c**, $ZrO_2$ dielectrics. Changes in exciton and trion emission indicate modulation of carrier density by gate WF. **d**, Extracted exciton-to-trion intensity ratio as a function of gate WF, extracted from multiple measurement points across different dielectrics. Inset: Schematic illustration of exciton and trion binding states, where trions form through Coulomb binding of an exciton with an additional electron. **e**, Schematic of the gate-dependent PL device used for dynamic carrier modulation studies. **f-g**, 2D colour maps of gate-dependent PL for $MoS_2$ on $ZrO_2$ dielectric with **f**, low-work-function Al gate and **g**, high-work-function Pt gate.

Photoluminescence (PL) spectroscopy measurements of monolayer $MoS_2$ were carried out to assess doping levels induced by different gate metals by comparing the relative intensities of neutral exciton (A) and negatively charged trion (A$^-$) peaks.[36, 37] The PL spectra for $MoS_2$ with hBN dielectric and different metal gates are plotted in **Figures**

**2a and S4**. A clear transition is observed from trion-dominant emission with Al gate to exciton-dominant peak with Pt gate. For MoS$_2$ on HfO$_2$ (**Figure 2b**), the PL spectra remain trion-dominant across all gate metals, showing negligible variation with gate WFs. These results indicates that interfacial interactions between HfO$_2$ and monolayer MoS$_2$ lead to substantial electron doping, rendering the channel resistant to depletion – even with high work function gate. In contrast, we observed a pronounced tunability in the exciton-to-trion ratio, similar to that on hBN, with high-$k$ ZrO$_2$ as the dielectric (**Figure 2c**). **Figure 2d** presents the exciton-to-trion ratio, an optical indicator of electron density in monolayer MoS$_2$, as a function of gate work function for the three different dielectric substrates. Effective WF-induced carrier modulation is evident only in the hBN and ZrO$_2$ systems, consistent with the device characteristics.

Gate-dependent PL measurements were conducted to further validate the influence of electrostatic modulation on exciton/trion dynamics. The device configuration used for these measurements is shown in **Figure 2e**. The gate tunability of the PL with ZrO$_2$ dielectric and Pt and Al gates is illustrated by **Figures 2f** and **2g**, where the PL emission evolves from trion-dominant (n-type) to exciton-dominant (neutral) as the gate-induced electron density decreases. The voltage at which this transition occurs is positive for Pt-gated devices and negative for Al-gated devices, indicating that MoS$_2$ channel is near charge neutrality under zero bias for Pt gate, but remains electron-doped for Al gate. The voltage difference at which the transition occurs between Pt gate and Al gate devices is ~ 0.8 V (details in **Figures S5** and **S6**). A similar gate-dependent trend is also observed in devices using hBN as the dielectric, as presented in **Figures S7** and **S8**.

To directly quantify the Fermi level modulation of monolayer MoS$_2$ induced by different metal gates, we performed synchrotron X-ray Photoelectron Spectroscopy (XPS) on devices with the three dielectric substrates. Specifically, we measured the binding energy of the Mo 3$d$, which reflects the energy difference between the Fermi level and the Mo 3$d$ core level (as illustrated in the inset of **Figure 3d**) [38]. As the work function of the metal gate increases, a systematic drop in the Mo 3$d$ binding energy is observed across all three dielectrics, indicating absence of electron doping in monolayer MoS$_2$ (**Figures 3a-c**). However, the extent of this Fermi level modulation varies depending on the dielectrics, as summarized in **Figure 3d**. It is evident that ZrO$_2$ enables the most effective gate control, where an increase in gate metal work function by ~1.1 eV leads to a downward Fermi level shift of ~ 0.6 eV in MoS$_2$. In contrast, the modulation achieved on HfO$_2$ is significantly weaker. With identical gate metal configurations, MoS$_2$ on HfO$_2$ consistently exhibits higher binding energy compared to those on ZrO$_2$ and hBN, indicating a pronounced electron-doping effect. Notably, even with a high work function Pt gate, the Fermi level position of MoS$_2$ on HfO$_2$ remains relatively high compared to Al-gated MoS$_2$ on hBN or ZrO$_2$, suggesting that effective electron depletion is challenging on HfO$_2$, which is consistent with the observations in PL measurements and FET characteristics. The energy difference between the Fermi level and the valence band maximum extracted by linear extrapolation of the valence band edge spectra are illustrated in **Figure S9**. Based on this energy difference, the carrier concentration is estimated using the formula: $n = N_c \ln\left[1 + \exp\left(\frac{E_F - E_i}{kT}\right)\right]$ (see supplementary information for details).[39, 40] The analysis reveals tunability of up to ~10$^{13}$ cm$^{-2}$ carrier concentration with gate metal in ZrO$_2$-based devices and up to ~ 10$^{12}$ cm$^{-2}$ in devices using hBN dielectrics. These values are consistent with the electrical measurements and gate-dependent PL result summarized in **Table S1**.

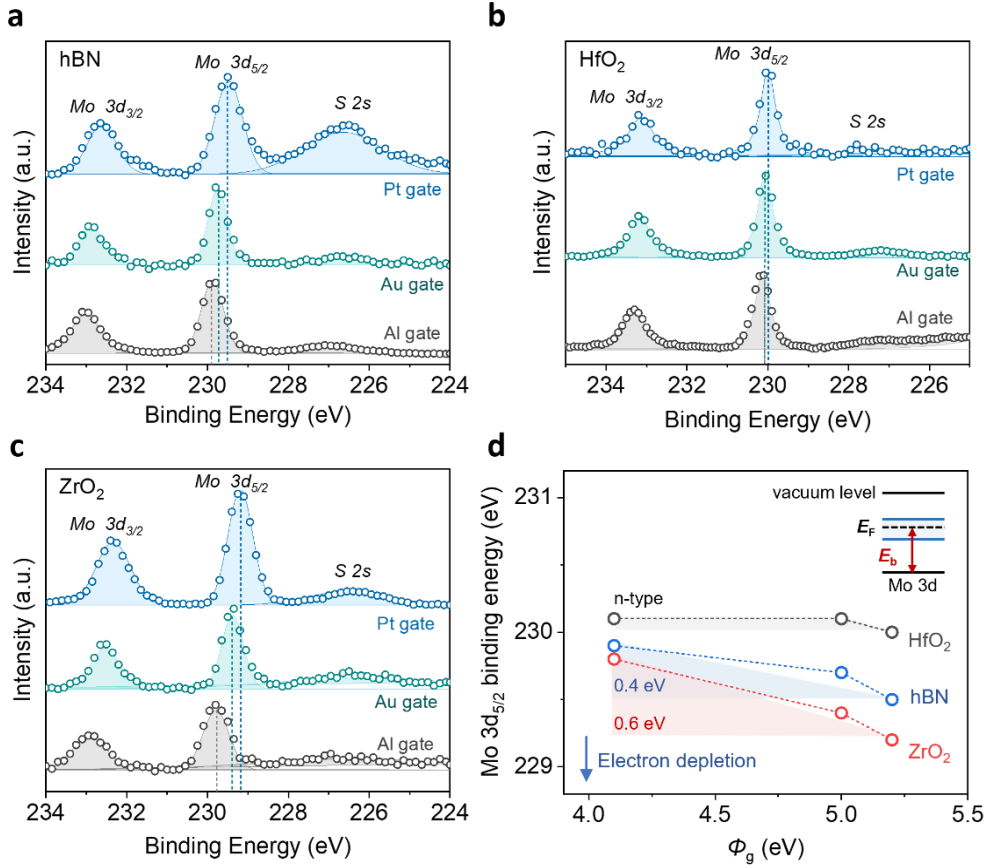

**Figure 3: Synchrotron XPS analysis of Fermi level modulation in monolayer MoS$_2$.** **a–c**, High-resolution XPS spectra of monolayer MoS$_2$ measured with different back gate metals buried under **a**, hBN, **b**, HfO$_2$, and **c**, ZrO$_2$ dielectrics. **d**, Extracted binding energy of the Mo 3$d_{5/2}$ peak as a function of gate work function. The Mo 3$d_{5/2}$ binding energy shifts indicate the relative position of the Fermi level with respect to the conduction band.

We further investigated key factors in V$_{th}$ modulation for transistors based on monolayer MoS$_2$. In MOSFETs, the V$_{th}$ can be expressed as:

$$V_{th} = 2\Phi_F + \gamma\sqrt{2\Phi_F} + V_{FB}$$

Where $\Phi_F$ is the Fermi potential, $\gamma$ is referred to as the body-effect coefficient and V$_{FB}$ is the flat band voltage. To estimate V$_{FB}$, we fabricated metal-oxide-semiconductor capacitors[41] (MOSCAP) using mechanically exfoliated few-layer MoS$_2$ – the device structure is shown in **Figure 4a**. High-frequency capacitance-voltage (C-V) characteristics (**Figure 4b**) reveal the expected transition from depletion to accumulation. In the depletion regime, the effective capacitance is dominated by the series combination of MoS$_2$ and ZrO$_2$, whereas in the accumulation regime, the ZrO$_2$ capacitance becomes dominant. Detailed analysis of capacitance evolution is provided in **Figure S10**. Importantly, a clear positive shift in the C-V curves is observed when varying the gate metals. The extracted V$_{FB}$, determined from the inflection points,[42] change from –1.2 V to 0.1 V, which is in good agreement with the V$_{th}$ extracted from FET measurements. This consistency confirms that ZrO$_2$ enables effective gate work function modulation with minimal interface or oxide state trapping. In contrast, MOSCAPs with HfO$_2$ exhibit fixed V$_{FB}$ values around –0.8 V, independent of the gate metal (**Figures S11-12**). More data points are provided in **Figure S13** to confirm the reproducibility.

Since V$_{FB}$ arises from both the work function difference and the interface/oxide charge, according to

$$V_{FB} = q(\Phi_g - \Phi_s) - \frac{Q_{ox}}{C_{ox}}$$

Where $\Phi_g - \Phi_s$ is the work function difference between metal gate and the semiconductor channel, $Q_{ox}$ is oxide charge density and $C_{ox}$ is the oxide capacitance. We further fabricated FETs with varying dielectric thicknesses to decouple the effect of Q$_{ox}$. For ZrO$_2$-based devices, the transfer curves and extracted V$_{th}$ remain nearly independent of oxide thickness, demonstrating that Q$_{ox}$ is negligible and that the V$_{th}$ modulation arises almost mainly from the gate-WF difference. In contrast, HfO$_2$-based devices demonstrated a negative shift in V$_{th}$ with dielectric thickness, highlighting significant contribution from trapped charges from interface states (**Figures 4f and S15**). Together, these results demonstrate that ZrO$_2$ enables reliable gate-work-function-driven V$_{th}$ engineering, whereas HfO$_2$ exhibits interface-state-dominated behavior that suppress effective V$_{th}$ modulation.

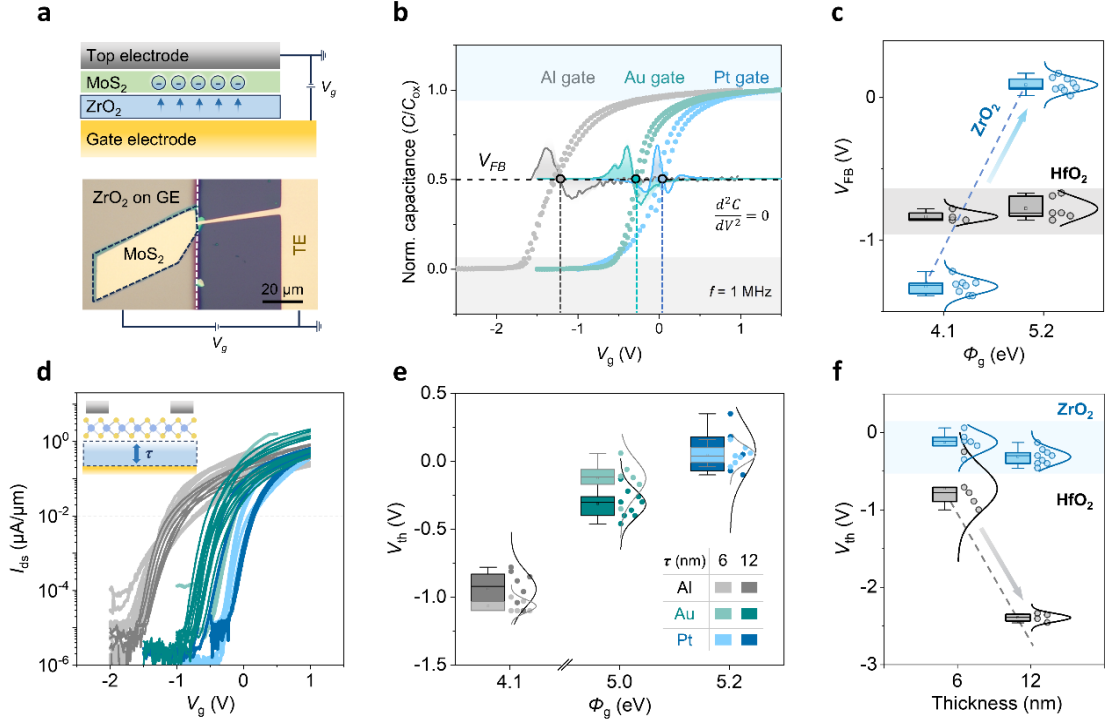

**Figure 4: Flat-band voltage extraction and dielectric-thickness-dependent $V_{th}$ modulation in $ZrO_2$ and $HfO_2$ devices. a**, Schematic and OM image of a representative MOSCAP device. **b**, Normalized high-frequency C-V measurements for $ZrO_2$-based MOSCAPs with varying bottom metal gates, along with the corresponding second-derivative curves used to extract $V_{FB}$ at the infection point. **c**, Extracted $V_{FB}$ values from multiple devices with Al and Pt gate metals on $ZrO_2$ and $HfO_2$. **d**, Transfer curves with different $ZrO_2$ thicknesses (6 nm, lighter curves; 12 nm, darker curves). **e**, Extracted $V_{th}$ for the corresponding devices in (**d**). **f**, $V_{th}$ as a function of dielectric thickness for $ZrO_2$ and $HfO_2$, using Au-gated devices as an example.

**Conclusion:**

In summary, we demonstrate modulation of threshold voltage in monolayer $MoS_2$ FETs using gate metals with different work functions. Our results highlight that this approach is only effective if the dielectric/semiconductor interface is mostly free of defects. Among the three dielectrics, $ZrO_2$ was found to provide a clean and well-coupled interface that allows tuning of carrier concentration in monolayer $MoS_2$ with gate metal work function variation. As a result, $ZrO_2$-based FETs exhibit linear $V_{th}$ modulation with gate metal work function, near-ideal subthreshold swing, and enhancement-mode operation when paired with high-work-function gate metals such as Pt. These findings provide design guidelines for monolayer MoS$_2$-based FETs that operate in enhancement mode.

**Experimental Section**:

**Sample preparation:** Monolayer MoS$_2$ was synthesized through chemical vapor deposition (CVD) using MoO$_3$ and sulfur as precursors. Specifically, 60 mg of sulfur powder was placed in an alumina boat located at the upstream edge of the furnace. A separate alumina boat containing 5 mg of MoO$_3$ was positioned at the center of the tube. Prior to growth, SiO$_2$/Si substrates were spin-coated with 0.5 mg/mL NaOH solution, serving as the growth promoter. The growth temperature was initially raised from room temperature to 120 °C over 3 mins and held for 20 mins to purge residual gases. Subsequently, the system was ramped to the growth temperature of 720 °C, while the MoS$_2$ growth was carried out for 12 mins. After growth, the sulfur source was promptly removed from the hot zone, and the furnace lid was opened to allow rapid cooling in ambient air. During the temperature ramp-up and cooldown (T < 650 °C), a nitrogen flow of 460 sccm was maintained to ensure an inert atmosphere, while during the high temperature growth phase (T ⩾ 650 °C), the flow rate was reduced to 60 sccm to optimize reaction kinetics.

**Dielectrics**: hBN flakes were mechanically exfoliated from bulk single crystals (NIMS, Japan). To minimize surface residue, exfoliation was performed using low-adhesion Ultron blue tape followed by transfer with home-made PDMS stamp. The thicknesses of selected flakes were measured by AFM and found to be approximately 12 nm. ZrO$_2$ films were deposited using E-beam evaporation (Nexdep, Angstrom Engineering). The chamber base pressure was maintained at 5×10$^{-6}$ Torr, and the deposition rate was controlled at 0.1 Å/s to ensure uniform film quality. Characterizations of the ZrO$_2$ dielectrics are given in **Figure S15**. HfO$_2$ films were deposited through atomic layer deposition (Fiji G2, Veeco). The whole process includes 60 cycles, yielding a film thickness of 6 nm.

**Device fabrications:**

Bottom gate electrodes (Al, Au, and Pt) were patterned by standard photolithography, followed by the deposition of 3 nm Ti adhesion layer and 12 nm of metal film using E-beam evaporation. After gate patterning, the dielectric layers were either transferred (for hBN) or deposited (for $ZrO_2$ and $HfO_2$) on top of the gate electrodes. Monolayer $MoS_2$ flakes were picked up and transferred by PDMS stamps method [43]. Following transfer, the samples were spin-coated with a double layer resist of MMA/PMMA. Source and drain electrodes were patterned with standard electron-beam lithography (EBL), followed by deposition of In/Au (6/60 nm) to avoid damage on monolayer channels and ensure low-contact resistance.

**Measurements:**

PL spectra were acquired using a confocal Raman/PL system (LabRAM Odyssey, Horiba) under the ×100 objective lens with the incident laser of 532 nm at 52 μW of incident power, which is sufficiently low to avoid photothermal damage to the monolayer samples. Gate-dependent PL measurements were performed by applying voltage bias using a source meter (Keithley 2450). During these measurements, the PL signals were collected using the ×50 objective lens, and the excitation power was set to 0.45 mW. All measurements were conducted under ambient conditions.

Synchrotron X-ray photoelectron spectroscopy measurements were conducted at the Diamond Light Source (UK), Beamline I09. The incident X-ray beam spot size was approximately 15 μm × 35 μm. Photon energies ranged from soft X-ray (1000 eV) to hard X-ray (3000 eV). All binding energies were calibrated against the standard C 1*s* peak (284.8 eV) to ensure consistency. The samples consisted of CVD-grown monolayer $MoS_2$ transferred onto various dielectric/gate stacks. To eliminate charging effects and ensure accurate energy referencing, all flakes were electrically grounded to the conductive sample holders during measurements.

Electrical measurements were performed with a Keithley 4200 semiconductor parameter analyzer. Prior to measurements, the chamber was pumped down to $10^{-6}$ Torr and maintained under vacuum for over 1 hour to minimize the influence of surface adsorbates and ensure measurement stability.

**FET performance calculation**

The $V_{th}$ was extracted at the drain current of 10 nA/μm from transfer curves.

The SS was extracted from the slope of logarithmic transfer characteristics:

$$SS = \frac{dV_g}{dlogI_d}$$

**Acknowledgements**

Y.W. and M.C. received funding from the European Research Council (ERC) under the European Union's Horizon 2020 research and innovation programme (grant agreement GA 101019828-2D-LOTTO, GA101220249-ATOMS), EPSRC (EP/T026200/1, EP/T001038/1 and EP/Z535680/1), Department of Science, Innovation and Technology and the Royal Academy of Engineering under the Chair in Emerging Technologies programme. Leyi Loh acknowledges support from the Royal Society Newton Fellowship (Grant No. NIF/R1/242837). K.K.P. acknowledge funding from Horizon Europe UK Research and Innovation (UKRI) Underwrite MSCA (EP/Y028287/1). K.W. and T.T. acknowledge support from the JSPS KAKENHI (Grant Numbers 21H05233 and 23H02052), the CREST (JPMJCR24A5), JST, and World Premier International Research Center Initiative (WPI), MEXT, Japan. We acknowledge Diamond Light Source for beamline under proposals SI38012-1, SI39914-2 and SI38086-1.

**Conflict of Interest**

The authors declare no conflict of interest.

**Author contribution**

Y.W. and M.C. conceived the project. Lixin Liu and Y.W. designed and performed the majority of experiments. H.Y. prepared the ALD samples and assisted with analyze the electrical characterization. Leyi Loh contributed to PL measurements and analysis. S.S. and K.K.P. assisted with gate-dependent PL experiments. Lixin Liu, H.Y., Y.W., and D.P. carried out the synchrotron XPS measurements with support from T.L. Lixin Liu worked on the figures with input from all co-authors. Lixin Liu, Y.W., and M.C. wrote the manuscript with contributions and feedback from all authors.

**Keywords**

2D semiconductor, threshold voltage, interface, high-*k* dielectric